\documentstyle[preprint,aps,eqsecnum,epsf,floats]{revtex} 
\tighten
\input epsf
\def\MeV{{\rm MeV}}
\def\GeV{{\rm GeV}}
\def\Tr{{\rm Tr\,}}
\def\nrcpt{NR\raise.4ex\hbox{$\chi$}PT\ }
\def\ket#1{\vert#1\rangle}
\def\bra#1{\langle#1\vert}
\def\ltap{\ \raise.3ex\hbox{$<$\kern-.75em\lower1ex\hbox{$\sim$}}\ }
\def\gtap{\ \raise.3ex\hbox{$>$\kern-.75em\lower1ex\hbox{$\sim$}}\ }

\def\CA{{\cal A}}

\def\CL{{\cal L}}

\def\pds{{\it PDS}\ }
\def\ms{MS}

\def\bfq{{\bf q}}

\def\frac#1#2{{\textstyle{#1\over#2}}}
\def\darr#1{\raise1.5ex\hbox{$\leftrightarrow$}\mkern-16.5mu #1}
\def\){\right)}
\def\({\left(}
\def\]{\right]}
\def\[{\left[}
\def\si{{}^1\kern-.14em S_0}
\def\siii{{}^3\kern-.14em S_1}
\def\diii{{}^3\kern-.14em D_1}
\def\fm{{\rm\ fm}}
\def\MeV{{\rm\ MeV}}
\def\CA{{\cal A}}
\def\Czzm{ {\cal A}_{-1[00]} }
\def\Cttm{{\cal A}_{-1[22]} }
\def\Ctzm{{\cal A}_{-1[20]} }
\def\Cztm{ {\cal A}_{-1[02]} }
\def\Czzz{{\cal A}_{0[00]} }
\def\Cttz{ {\cal A}_{0[22]} }
\def\Ctzz{{\cal A}_{0[20]} }
\def\Cztz{{\cal A}_{0[02]} }

\newcommand{\eqn}[1]{\label{eq:#1}}
\newcommand{\refeq}[1]{(\ref{eq:#1})}
\newcommand{\eq}{eq.~\refeq}
\newcommand{\eqs}[2]{eqs.~(\ref{eq:#1}-\ref{eq:#2})}
\newcommand{\eqsii}[2]{eqs.~(\ref{eq:#1}, \ref{eq:#2})}

\newcommand{\beq}{\begin{eqnarray}}
\newcommand{\eeq}{\end{eqnarray}}

\begin{document}

\preprint{\vbox{
\hbox{ DOE/ER/40561-357-INT98-00-5}
\hbox{ NT@UW-98-08}
\hbox{ CALT-68-2161} }}
\bigskip
\bigskip
\title{Two-nucleon systems from effective field theory}
\author{David B. Kaplan}  \address{ Institute for Nuclear Theory,
University of Washington, Seattle, WA 98915   \\  {\tt
dbkaplan@phys.washington.edu} }  \author{Martin  J. Savage}  \address{
Department of Physics, University of Washington,  Seattle, WA 98915
\\ {\tt savage@phys.washington.edu} }  \author{Mark B. Wise} \address{
California Institute of Technology, Pasadena, CA 91125  \\  {\tt
wise@theory.caltech.edu} }  \maketitle
\begin{abstract}
We elaborate on a new technique for computing properties of nucleon-nucleon
interactions
in terms of an effective field theory derived from low energy $NN$
scattering data. 
Details of
how the expansion is carried out to higher orders 
are presented.   
Analytic formulae are given for the amplitude to 
subleading order in both the $\si$ and $\siii-\diii$ channels. 

\end{abstract}

\vskip 2in

%
%
%
%
\vfill\eject

\section{Introduction}
\label{sec:1}

Effective field theory appears to be an ideal tool for the study
of low energy nuclear physics, as the nucleon energies are typically well
below the complex spectrum of hadrons that exist with masses greater
than about $1\GeV$.
An example of the successful application of effective
field theory to low energy hadronic physics is chiral perturbation theory, 
which exploits the fact that the lightest  pseudoscalar mesons are
approximate Goldstone bosons (for a recent review, see \cite{ManRev}).  
Even though an analytic description of pions in terms
of quarks and gluons is impossible, our ignorance can be parametrized
in an effective theory such that a perturbative calculation of  pion
interactions involving only a few parameters agrees well with
experiment.  Central to the utility of chiral perturbation theory for
mesons is that  there is  a clear
power counting scheme, so that one can include all effects to a given
order, and estimate the size of errors incurred in the approximation.

Recently we applied the same procedure to nucleon-nucleon
interactions,  outlining a method to consistently
expand the $NN$ interaction in powers of $p$ and $m_\pi$, where $p$
is the momentum of each nucleon in the center of mass frame, and 
$m_\pi$ is the pion mass  \cite{KSWb}.  The analysis was inspired by 
Weinberg's proposal \cite{Weinberg1}\ that effective field theory
could be profitably used in nuclear physics, as well as by subsequent work
\cite{KoMany,Parka,KSWa,CoKoM,DBK,cohena,Fria,Sa96,LMa,GPLa,Adhik,RBMa,Bvk,Parkb}.
The original idea was to exploit the approximate chiral symmetry of the strong
interactions, which gives rise to a hierarchy of length scales
 between the Compton wavelengths of the
vector mesons and the pions.  
In an effective field theory, 
short distance nucleon-nucleon interactions are
encoded in a derivative expansion of local operators. This is in
contrast with the various models of extended nucleon-nucleon  potentials
 with  free parameters  chosen to fit scattering data.  
These models can fit the
nucleon-nucleon phase shifts to great accuracy, but suffer from
several deficiencies not shared by the effective field theory
approach:  they are not useful for computing inelastic processes, they
give no insight into three-nucleon forces, and they are numerically
intensive to use in the $N$-body problem.  Furthermore, there is no
systematic way of anticipating the errors one should expect when using
these potentials.  Another advantage of effective field theory is
that it can easily incorporate chiral symmetry, and
 can be naturally extended to discuss
systems with strange quarks, such as hypernuclei \cite{SavageWise} and
kaon condensation \cite{KaplanNelson,kcon}.

However, the effective field theory analysis of the two-nucleon system
is complicated by the
existence of other length scales, 
in particular the $S$-wave 
scattering lengths, which are many times longer than the 
pion Compton wavelength. The existence of large scattering lengths implies
 that the underlying physics
at short distance is both nonperturbative and ``finely tuned'', in the 
sense that the interactions must be near a critical value. In Weinberg's 
original work 
\cite{Weinberg1}\  a power counting scheme was proposed that involved
summing a particular infinite class of Feynman graphs at each order in the expansion. 
However, it was shown in refs. \cite{KSWb,KSWa} that graphs at a given order 
in that expansion require counterterms corresponding to operators treated 
as being higher order, and that therefore the  expansion proposed in \cite{Weinberg1}\  
is inconsistent.
This was also demonstrated in models where a perturbative matching could
be performed \cite{LMa}.

Ref. \cite{KSWb}\ 
presented a different expansion that cures this problem, and applied it 
to $NN$ scattering in  the $\si$ and $\siii-\diii$ channels. 
In this paper we begin
by giving further details about the expansion, analyzing in detail 
a model of heavy bosons interacting via meson exchange.
This simplified
theory serves to explain the special treatment accorded systems with
large scattering lengths, and explains the virtues of the \pds subtraction 
scheme and renormalization group analysis 
introduced in ref. \cite{KSWb}.  We then explain how the power counting
is extended to the theory with pions
and give the explicit analytic formulae for the spin-triplet
scattering amplitudes to subleading order.

\section{Effective field theory for nonrelativistic scattering: a toy example}
\label{sec:2}

In this section we present a toy model of heavy spinless ``nucleons'' $\tilde N$ 
interacting via a Yukawa interaction characterized by a scale $\Lambda$. We 
then construct the effective field theory describing scattering at
 momenta $p\ll \Lambda$,  consisting entirely of 
contact interactions in a derivative expansion.
 Since these local operators are singular, this formulation of the low energy 
theory  necessarily introduces divergences that must be dealt with by the 
conventional regularization and renormalization procedures, so that the final 
result is independent of a momentum cutoff. We show how to organize the Feynman 
graphs in the effective theory in a consistent power counting scheme so that the 
scattering amplitude can be expanded in powers of $p/\Lambda$. Since 
the sizes of all the coupling constants in the effective theory depend on the 
subtraction scheme used to render diagrams finite, the development of the power 
counting scheme is intimately related to the renormalization procedure used.  We 
show why the \pds  subtraction scheme introduced in \cite{KSWb}\ is particularly 
well suited for this problem,  and we show how a renormalization group analysis 
is useful in the case of systems with large scattering length, such as those seen 
in the realistic problem of nucleon-nucleon scattering.

 It should be no surprise that the effective field theory expansion for the 
toy system is simply related to the conventional effective range expansion,
 and so the machinery of quantum field theory may appear to be heavy handed 
and superfluous. Nevertheless,  the field theoretic language that we develop 
in this section is readily extended to the realistic problem of interest: 
nucleons  interacting via both short range interactions and long range pion 
exchange.  In the realistic problem, effective field theory is not equivalent 
to an effective range expansion, and is the only framework that can consistently incorporate chiral 
symmetry and relativistic effects without resorting to 
phenomenological  models.

We assume that the spinless bosons $\tilde N$ are nonrelativistic with  mass $M$,
carry a conserved charge (``baryon
number''), and interact via the exchange of a meson $\phi$ with mass $\Lambda$
and coupling $g$.  At tree level, meson exchange gives rise to the Yukawa
interaction
\beq
V(r)= -{g^2\over 4\pi} {e^{-\Lambda r}\over r},
\eeq
and the Schr\"odinger equation for this system may be written as
\beq
\left[ -\nabla_x^2 + \eta {e^{-x}\over x} - {p^2\over \Lambda^2}\right]\Psi=0\
,\\
 \vec x\equiv \Lambda \vec r, \qquad \eta\equiv{g^2 M\over 4\pi \Lambda}, 
\qquad p^2\equiv M E\ .
\eqn{sequ}
\eeq
Note that $p$ is the magnitude of the momentum carried by each $\tilde N$ 
particle in the center of mass frame.
Evidently there are two options for a perturbative solution for the $S$-matrix
for this system.  The first is an expansion in powers of $\eta$, the familiar Born
expansion.  
An alternative is to expand in powers of $p/\Lambda$, which is
the expansion parameter used in effective field theory.   An important feature
of the low energy expansion is that it can provide accurate results in terms 
of a few phenomenological parameters even for nonperturbative $\eta$.  This is
the regime we are interested in, and so we will assume throughout that $\eta\sim 1$.

The quantity that is natural to calculate in a field theory is the sum of Feynman 
graphs, which gives the amplitude $i\CA$, related to the $S$-matrix by
\beq
S= 1 + i {Mp\over 2\pi}\CA\ .
\eeq
For $S$-wave scattering, $\CA$ is related to the phase shift $\delta$ 
through the relation
\beq
\CA = {4\pi \over M}{1\over p\cot\delta -i p}\ .
\eqn{amp}
\eeq
From quantum mechanics it is well known that it is not $\CA$, but rather the quantity 
$p\cot\delta$, which has a nice momentum
expansion for $p\ll\Lambda$ (the effective range expansion):
\beq
p\cot\delta = -{1\over a} + {1\over 2}\Lambda^2\sum_{n=0}^\infty {r}_n
\({p^2\over \Lambda^2}\)^{n+1}\ ,
\eqn{erexp}
\eeq
where $a$ is the scattering length, and 
$r_0$ is the effective range.
So long as $\eta\sim 1$, the case we will be interested in, 
the coefficients  $r_n$ are generally $O(1/\Lambda)$ for all $n$, but $ a$ can 
take on any value, diverging as $\eta$ approaches one of the critical couplings 
$\eta_k$ for which there is a boundstate at threshold. (The lowest critical coupling 
is found numerically to be $\eta_1=1.7$.) Therefore the radius of convergence of a  
momentum expansion of $\CA$ depends on the size
of the scattering length $a$.  In
the next section the situation where the scattering length is of natural
size $|a|\sim 1/\Lambda$ is considered,
while in  the subsequent section we discuss the case
$|a|\gg 1/\Lambda$, which is   relevant for realistic $NN$ scattering.

\subsection{The momentum expansion for a scattering length of natural size}
\label{sec:2a}

In the regime $|a|\sim  1/\Lambda$ and $|r_n|\sim  1/\Lambda$,
$\CA$ has a simple momentum expansion in terms of
the low energy scattering data,
\beq
\CA = -{4\pi a \over M }\[ 1 - i a p +(ar_0/2-a^2) p^2 + O(p^3/\Lambda^3)\]\ ,
\eqn{aexp}
\eeq
which converges up to momenta $p\sim \Lambda$.   It is this expansion that we
wish to reproduce in an effective field theory.

The effective field theory of $\tilde N$ particles 
interacting  through contact interactions has the following  Lagrangian:
\beq
 \CL=
{{\tilde N}^\dagger} \left( i\partial_t + \nabla^2/2M\right) \tilde N
+(\mu/2)^{4-D}\left[C_0 ({{\tilde N}^\dagger} \tilde N)^2
+ C^{(1)}_2 ({{\tilde N}^\dagger} {  \mathop\nabla^\leftrightarrow} \tilde N)^2
+  C^{(2)}_2  [i\vec\nabla({{\tilde N}^\dagger } \tilde N)]^2 +...\right]\ .
\eqn{yukeff}
\eeq
The sum of Feynman diagrams computed in this theory gives us the 
amplitude $\CA$.
As we will be using dimensional regularization for the loop integrals in this theory, 
the spacetime dimension is given by $D$. 
Dimensional regularization is the preferred regularization scheme 
as it preserves gauge symmetry and chiral symmetry, 
as well as Galilean invariance (or Lorentz invariance, for relativistic systems).   
The ellipses indicates higher derivative operators,  
and $(\mu/2)$ is an
arbitrary mass scale introduced to allow the couplings $C_{2n}$
multiplying operators containing $\nabla^{2n}$  to have the same dimension for
any $D$.  We focus on  the $S$-wave channel
(generalization to higher partial waves is straightforward), and assume
that $M$ is very large so that relativistic effects can be ignored.  
The tree level $S$-wave amplitude is
\beq
i \CA_{\rm tree} = -i(\mu/2)^{4-D} \sum_{n=0}^\infty C_{2n}(\mu) p^{2n}\ ,
\eqn{tree}
\eeq
where the coefficients $C_{2n}(\mu)$ are various linear combinations of the
couplings in the Lagrangian \eq{yukeff} contributing to $S$-wave scattering.

\begin{figure}[t]
\centerline{\epsfysize=1 in \epsfbox{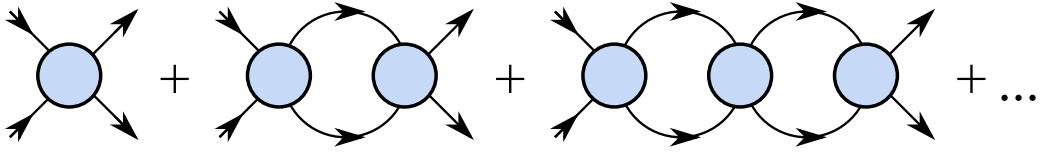}}
\noindent
\caption{\it The bubble chain arising from local operators.}
\label{bubbles}
\vskip .2in
\end{figure}

The loop integrals
one encounters in diagrams shown in Fig.~\ref{bubbles}
are of the form
\beq
\openup3\jot
I_n&\equiv& -i(\mu/2)^{4-D} \int {{\rm d}^D q\over (2\pi)^D}\, {\bf q}^{2n}
\({i\over E/2 + q_0 -{\bf q}^2/ 2M + i\epsilon}\)\({i\over E/2 - q_0
-{\bf q}^2/2M + i\epsilon} \) 
\nonumber\\
&=& (\mu/2)^{4-D} \int {{{\rm d}}^{(D-1)}{\bf  q}\over (2\pi)^{(D-1)}}\, 
{\bf q}^{2n} \({1\over E  -{\bf q}^2/M + i\epsilon}\) 
\nonumber\\
&=& -M (ME)^n (-ME-i\epsilon)^{(D-3)/2 } \Gamma\({3-D\over 2}\)
{(\mu/2)^{4-D}\over  (4\pi)^{(D-1)/2}}\ .
\eqn{loopi}
\eeq
In order to define the theory, one must specify a subtraction scheme; different subtraction schemes amount to a reshuffling between contributions from
the vertices and contributions from the the UV part of the 
loop integration.  
For the case we are considering, $|a|, |r_n|\sim 1/\Lambda$, 
it is convenient to
use the minimal subtraction scheme ($\ms$) which amounts to subtracting any
$1/(D-4)$ pole before taking the $D\to 4$ limit.  
The integral \eq{loopi} doesn't exhibit any such poles and so the result is simply
\beq
I_n^{\ms}=  (ME)^n \({M\over 4\pi}\)\sqrt{-ME-i\epsilon}=
-i\({M\over 4\pi}\) p^{2n+1}  \ .
\eqn{inval}
\eeq
A nice feature of this scheme is that the factors of $q$ inside the loop get converted
to factors of $p$, the external momentum.  
Therefore one can use the on-shell,
tree level amplitude \eq{tree} as the internal  vertices in loop diagrams and 
summing the bubble diagrams gives
\begin{eqnarray}
\CA & = & -{  \sum C_{2n} p^{2n} 
\over
1 + i(M p/4\pi) \sum C_{2n} p^{2n} }
\ .
\end{eqnarray}
 Since there are no poles at $D=4$ in the $\ms$ scheme the coefficients $C_{2n}$ are 
independent of the subtraction
point $\mu$.
The power counting in the $\ms$ scheme is particularly simple: each vertex 
$C_{2n}\nabla^{2n}$ counts as order $p^{2n}$, while each loop brings in a factor of $p$.  
The amplitude may be expanded in powers of $p$ as
\beq
\CA=\sum_{n=0}^\infty \CA_n
\ \ \ ,\ \ \ 
\CA_n\sim O(p^n)
\eqn{ampexpand}
\eeq
where the $\CA_n$ each arise from graphs with $L\le n$ loops and can be equated to the low
energy scattering data \eq{aexp} in order to fit the $C_{2n}$ couplings.  In particular,
$\CA_0$ arises from the tree graph with $C_0$ at the vertex; $\CA_1$ is given by the 
1-loop diagram with two $C_0$ vertices; $\CA_2$ is gets contributions from both
 the 2-loop diagram with
three $C_0$ vertices, as well as the tree diagram with one $C_2$ vertex, and so forth.  
Thus the first three terms are
\beq
\CA_0= -C_0\ ,\qquad
\CA_1= i C_0^2{Mp\over 4\pi}\ ,\qquad
\CA_2=  C_0^3\({Mp\over 4\pi}\)^2-C_2p^2
\ .
\eqn{effexp}
\eeq
Comparing \eqsii{aexp}{effexp} we find for the first two couplings of the effective theory
\beq
C_0 = {4\pi a\over M}\ ,\qquad C_2 = C_0 {a r_0\over 2}
\ . 
\eqn{cfit}
\eeq
In general, when the scattering length has natural size,
\beq
C_{2n} \sim {4\pi \over M\Lambda} {1\over \Lambda^{2n}}
\ .
\eeq
Note that the effective field theory calculation in this scheme
is completely perturbative even though $\eta\sim 1$ and
there may be a boundstate well below threshold.  
The point is,
that when there are  no poles in $\CA$ in the region $|p|\ltap\Lambda$, the amplitude is amenable
to a Taylor expansion in $p/\Lambda$ in that region, and  with a suitable subtraction scheme this
Taylor expansion can correspond to a perturbative sum of Feynman graphs.

\subsection{The momentum expansion for large scattering length}
\label{sec:2b}

Now consider the case $|a|\gg 1/\Lambda$, $|r_n|\sim 1/\Lambda$, which 
is of relevance to realistic $NN$ scattering.
For a nonperturbative interaction ($\eta\sim 1$) with
a boundstate near threshold, the  expansion of $\CA$ in powers of $p$
is of little practical value, as it breaks down for momenta $p\gtap 1/|a|$,  
far below $\Lambda$.
In the above effective theory, this occurs because the couplings $C_{2n}$ 
are anomalously large, $C_{2n}\sim 4\pi a^{n+1} / M\Lambda^n$.
However, the problem is
not with the effective field theory method, 
but rather with the subtraction scheme chosen.

Instead of reproducing the expansion of the amplitude shown in \eq{aexp}, one needs 
to expand in powers of $p/\Lambda$ while retaining $ap$ to all orders:
\beq
\CA = -{4\pi\over M}{1\over (1/a + i p)}\[ 1 + {r_0/2 \over (1/a + ip)}p^2 +  
{(r_0/2)^2\over (1/a + ip)^2} p^4 + {(r_1/2\Lambda^2)\over (1/a + ip) } p^4 +\ldots\]
\eqn{aexp2}
\eeq
Note that  for $p>1/|a|$ the terms  in this expansion scale as $\{p^{-1}, p^0,p^1,\ldots\}$.
Therefore, the expansion in the effective theory should take the form
\beq
\CA=\sum_{n=-1}^\infty \CA_n
\ \ \ ,\ \ \ 
\CA_n\sim O(p^n)
\eqn{ampbiga}
\eeq
instead of the expansion in \eq{ampexpand}. 
Again, the task is to compute the $\CA_n$ in the 
effective theory, equate to the appropriate expression 
in \eq{aexp2}, and thereby fix the 
$C_{2n}$ coefficients.  For example, 
\beq
\CA_{-1} =  -{4\pi\over M}{1\over (1/a + i p)}\ .
\eqn{ami}
\eeq

 As we have seen in the previous section, any single diagram  computed in the effective theory 
is proportional to positive powers of $p$.  The leading term $\CA_{-1}$ must therefore involve 
summing an infinite set of diagrams. It is easy to see that the leading term in \eq{aexp2}  
can be reproduced by the sum of bubble diagrams with $C_0$ vertices \cite{Weinberg1}, 
which yields
\beq
{\cal A}_{-1} = { -C_0\over \left[1 + {C_0 M\over 4\pi} ip\right]}\ ,
\eeq
in the $\ms$ scheme.  
Comparing this with \eq{ami} we find $C_0=4\pi a/M$, as in the 
previous section.  However, there is no  expansion parameter that justifies this summation:  
each individual graph in the bubble sum goes as $(4\pi a/M)(i a p)^L$, where $L$ is the number 
of loops. Therefore each graph in the bubble sum is bigger than the preceding one, for $|ap|>1$, 
while they sum up to something small.  

This is an unacceptable situation for an effective field theory;  it is important to have an 
expansion parameter so that one can  identify the order of any particular graph, and sum them 
up consistently.
Without such an expansion parameter, one cannot determine the size of omitted contributions, 
and one can end up retaining certain graphs  while dropping operators
needed to renormalize those graphs.  This results 
in a model-dependent description of the short distance physics,
as opposed to a proper effective field theory calculation.

Since the sizes
of the contact interactions depend on the renormalization scheme one uses, the
task becomes one of identifying the appropriate subtraction scheme that makes the 
power counting simple and manifest.
The $\ms$ scheme fails on this point;  however  this is not a 
problem with dimensional regularization, as has been frequently suggested, but rather a problem with the
minimal subtraction scheme itself.  The momentum space subtraction at threshold used in ref.
\cite{Weinberg1}\ behaves similarly.  

An alternative regularization and renormalization
scheme is to use a momentum cutoff equal to $\Lambda$. 
Then for large $a$, $C_0 \sim (4\pi/M\Lambda)$, and each additional loop contributes 
a factor of  $C_0(\Lambda + ip)M/4\pi \sim (1+ip/\Lambda)$.
For $\Lambda\gg p$ the factor of $p/\Lambda$ is small relative to the $1$,
but neglecting it would fail to reproduce the desired result \eq{ami}.  
The problem is that 
there are significant cancellations between terms in this scheme.

Evidently, since $\CA_{-1}$ scales as $1/p$, the desired expansion 
would have each individual graph contributing to $\CA_{-1}$ scale as $1/p$.  
As the tree level contribution is $C_0$, we must have
$C_0$ be of size $\propto 1/p$, and each additional loop must be $O(1)$. 
This can be achieved by using  dimensional regularization and 
 the \pds (power divergence  subtraction)
scheme introduced in ref. \cite{KSWb}. 
The \pds scheme involves subtracting from the
dimensionally regulated loop integrals not only the $1/(D-4)$ poles corresponding
to log divergences, as in $\ms$, but also 
poles in lower dimension which correspond to power law divergences at $D=4$.  
The integral $I_n$ in 
 \eq{loopi}\ has a pole in $D=3$ dimensions which can be removed by adding to $I_n$ 
the counterterm
\beq
\delta I_n = -{M(ME)^n \mu\over 4\pi (D-3)},
\eeq
so that the subtracted integral in $D=4$ dimensions is
\beq
I_n^{PDS} = I_n + \delta I_n = - (ME)^n \left({M\over 4\pi}\right) (\mu + ip).
\eqn{ipds}
\eeq 
An alternative subtraction scheme with similar power counting  is to perform
a momentum subtraction at $p^2=-\mu^2$, as recently suggested in ref. \cite{Gegelia}.
The \pds scheme  retains the nice feature of $\ms$ that
powers of $q$ inside the loop integration are effectively replaced by powers 
of the external momentum $p$.
Note that at $\mu=0$, \pds and $\ms$ are the same.
In this subtraction scheme 
\begin{eqnarray}
\CA & = & -{  \sum C_{2n} p^{2n} 
\over
1 + M(\mu+ip)/4\pi \sum C_{2n} p^{2n} }
\ .
\eqn{answer}
\end{eqnarray}

\begin{figure}[t]
\centerline{\epsfysize=3 in \epsfbox{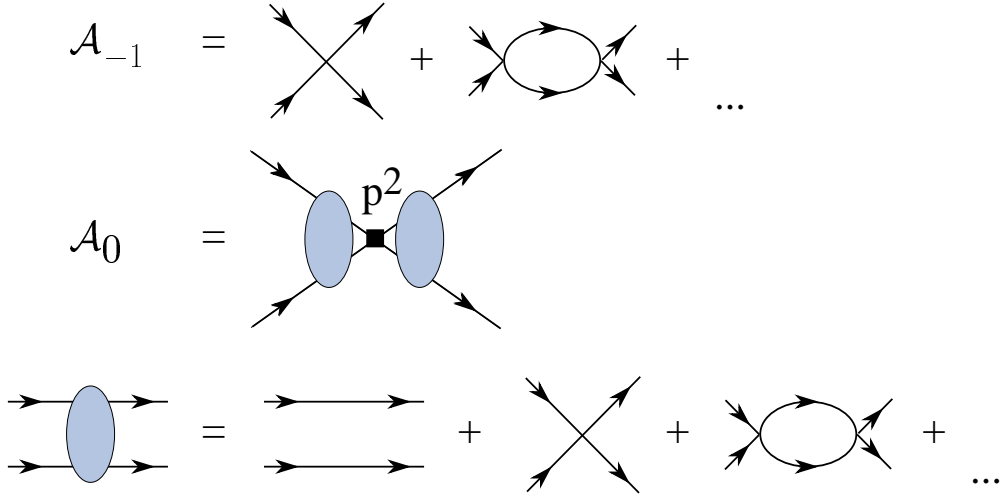}}
\noindent
\caption{\it Leading and subleading contributions arising from local operators.}
\label{FG1S0_m1}
\vskip .2in
\end{figure}

The amplitude $\CA$ is independent of the subtraction point $\mu$ and
this fact determines the $\mu$ dependence of the coefficients, $C_{2n}$. 
In the \pds scheme one finds that for $\mu\gg 1/|a|$, 
the couplings $C_{2n}(\mu)$ scale as
\beq
C_{2n}(\mu) \sim {4\pi \over M \Lambda^n \mu^{n+1}}\ ,
\eqn{cscale}
\eeq
so that if we take $\mu \sim p$, $C_{2n}(\mu)\sim 1/p^{n+1}$.  
A factor of $\nabla^{2n}$ at a vertex scales as $p^{2n}$, while each loop contributes a factor of 
$p$.  Therefore, the leading order contribution to the scattering amplitude $\CA_{-1}$  
scales as $p^{-1}$ and consists of the sum of bubble diagrams with $C_0$ vertices; 
contributions to the amplitude scaling as higher powers of $p$ come from perturbative 
insertions of derivative interactions, dressed to all orders by $C_0$.  The first three 
terms in the expansion are
\beq
{\cal A}_{-1}&=& { -C_0\over \left[1 + {C_0 M\over 4\pi} (\mu + ip)\right]}\ ,\nonumber\\
{\cal A}_0    &=& { -C_2 p^2\over \[1 + {C_0 M\over 4\pi}(\mu + ip)\]^2}\ ,\nonumber\\
{\cal A}_1    &=& \({ (C_2 p^2)^2M(\mu+ ip)/4\pi\over \[1 + {C_0 M\over 4\pi}(\mu + ip)\]^3}
-{ C_4 p^4\over \[1 + {C_0 M\over 4\pi}(\mu + ip)\]^2}\) \ ,
\eeq
where the first two correspond to the Feynman diagrams
in Fig.~\ref{FG1S0_m1}.

Comparing with the expansion of the amplitude \eq{aexp2}, these expressions  
relate the couplings $C_{2n}$ to the low energy scattering data $a$, $r_n$:
\beq
C_0(\mu) &=& {4\pi\over M }\({1\over -\mu+1/a}\)\ ,\nonumber\\
C_2(\mu) &=&  {4\pi\over M }\({1\over -\mu+1/a}\)^2 {{
r}_0\over 2}\ ,\nonumber\\
C_4(\mu) &=&  {4\pi\over M }\({1\over -\mu+1/a}\)^3 \[{1\over 4}
{ r}_0^2 + {1\over 2} {{ r}_1\over\Lambda^2} \({-\mu+1/a}\)\]\ .
\eqn{cvals}
\eeq
Note that assuming $r_n\sim 1/\Lambda$, these expressions are
consistent with the scaling law in \eq{cscale}. 

This power counting relies entirely on the behavior of
$C_{2n}(\mu)$ as a function of $\mu$ 
given in \eq{cscale}. The dependence of $C_{2n}(\mu)$ on
$\mu$ is determined by the requirement that the amplitude be independent of the
arbitrary parameter $\mu$.  The physical parameters $a$, $ r_n$ enter as
boundary conditions on the RG equations. 

The beta function for each of the couplings $C_{2n}$ is defined by 
\beq
\beta_{2n} \equiv \mu {{\rm d}C_{2n}\over {\rm d}\mu}\ ,
\eqn{betadef}
\eeq
and they can be computed by requiring that any physical quantity (e.g. the
scattering amplitude) be independent of $\mu$.
In the \pds scheme, the $\mu$ dependence of the $C_{2n}$ coefficients enters
either logarithmically or linearly, associated with simple $1/(D-4)$ or
$1/(D-3)$ poles respectively.  
The functions $\beta_{2n}$  follow straightforwardly from
$\mu{d\over d\mu}(1/\CA)=0$, using the expression for $\CA$ 
in \eq{answer}.  This gives
\beq
\beta_{2n} = {M\mu\over 4\pi} \sum_{m=0}^{n} C_{2m} C_{2(n-m)}\ .
\eqn{beta2n}
\eeq
\begin{figure}[t]
\centerline{\epsfysize=3 in \epsfbox{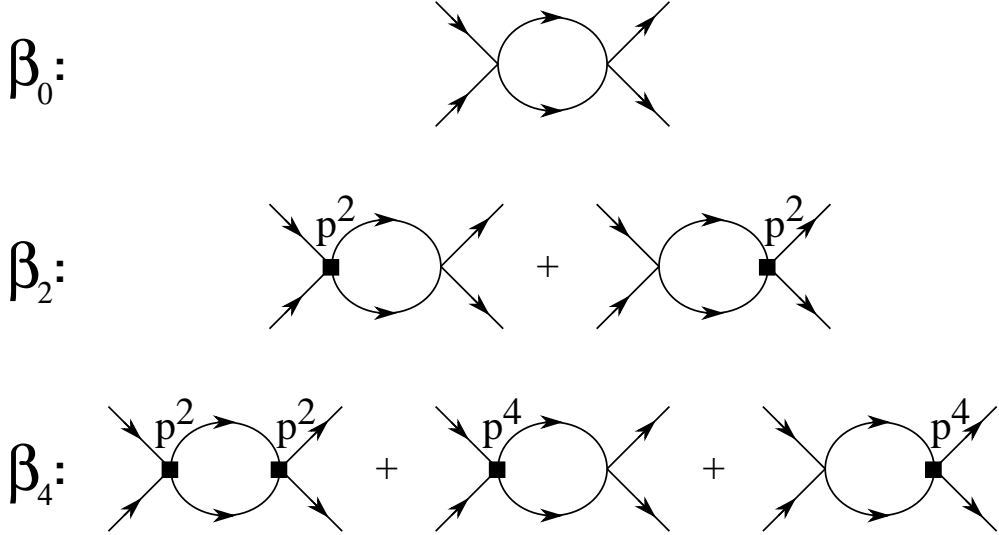}}
\noindent
\caption{\it Graphs contributing to the $\beta$-functions for $C_{2n}$}
\vskip .2in
\label{betanopi}
\end{figure}
These $\beta$-functions can also be computed from the one-loop diagrams 
shown in Fig.~\ref{betanopi}.

We examine the RG equations for the first two couplings, $C_0$ and $C_2$, 
in order to 
explicitly show how one recovers the results in \eq{cvals} from solving the
renormalization group equations. From \eq{beta2n} it follows that
\beq
\beta_0&=& {M\mu\over 4\pi} C_0^2\ ,\\
\beta_2&=& 2 {M\mu\over 4\pi} C_0 C_2\ .
\eqn{beta024}
\eeq
Integrating these equations relates the $C_{2n}$ coefficients at two different
renormalization scales $\mu$ and $\mu_0$. Comparing the theory with $\CA$ and its 
derivatives at $\mu=p=0$ determines the initial values $C_{2n}(0)$ as in \eq{cfit}.
The solution for $C_0(\mu)$ is
\beq
 C_0(\mu) = {C_0(\mu_0)\over 1+ C_0(\mu_0)M  (\mu_0 - \mu)/4\pi}
 \ \ \ .
\eqn{c0rg}
\eeq
With the boundary condition at $\mu_0=0$ provided by \eq{cfit},
 $C_0(0)=4\pi a/M$, 
we arrive at the result derived previously for $C_0(\mu)$ in \eq{cvals}:
\beq
C_0(\mu) = {4\pi\over M }\({1\over -\mu+1/a}\)\ .
\eqn{c0run}
\eeq
  
The RG equation for $C_2$ yields
\beq
C_2(\mu) = C_2(\mu_0)\({C_0(\mu)\over C_0(\mu_0)}\)^2\ ,
\eqn{c2rg}
\eeq
which when combined with the value of $C_0(0)$ and the boundary condition,
$ C_2(0) = C_0(0) a r_0/ 2$, 
from \eq{cfit}, yields for $C_2(\mu)$
\beq
C_2(\mu) = {4\pi\over M }\({1\over- \mu+1/a}\)^2 {{
r}_0\over 2}\ ,
\eeq
as we found previously in \eq{cvals}.

It is possible to solve the complete, coupled RG equation 
\beq
\mu {{\rm d}\ \over {\rm d}\mu}C_{2n} = {M\mu\over 4\pi}\sum_{m=0}^n C_{2m} C_{2(n-m)}
\eeq
for the leading small $\mu$ behavior of each of the coefficients $C_{2n}$ . The solution, for $n>0$ is
\beq
C_{2n}(\mu) = {4\pi \over M(-\mu+1/a)} \({r_0/2\over -\mu+1/a}\)^n + O(\mu^{-n})\ .
\eeq
First note that the scaling property in \eq{cscale} is realized: $C_{2n}(\mu)\propto \mu^{-(n+1)}$ for 
$\mu\gg |1/a|$.  What is curious is that this leading behavior does not entail a new 
integration constant for each $n$, but only depends on the two parameters $a$ and $r_0$ 
encountered when solving for $C_0(\mu)$ and $C_2(\mu)$;  this is due to a quasi-fixed 
point behavior of the RG equations ---  the $C_{2n}$ couplings are being driven primarily 
by lower dimensional interactions. One can see this explicitly in our formula \eq{cvals} 
for $C_4$, where the leading $O(\mu^{-3})$ part of $C_4$ depends only on $r_0$, while 
the subleading $O(\mu^{-2})$ part is proportional to $r_1$.

This behavior allows us to establish a connection between the present work, and the 
method of introducing an $s$-channel dibaryon discussed in ref. \cite{DBK}.  The leading 
$\mu$ behavior of all of the $C_{2n}$ coefficients is determined by the effective range $r_0$.  
If one resums this leading behavior at the $\tilde N \tilde N$ vertex one finds (for 
$\mu\sim p\gg 1/|a|$)
\beq
\sum_{n=0}^\infty C_{2n}(\mu) p^{2n} &=&  
{4\pi \over M}{1\over -\mu+1/a -{r_0 p^2\over 2}}+ O(p^2)\nonumber\\
&=&-{(8\pi /  M^2 r_0)\over E-(-\mu+1/a)/Mr_0}\ .
\eeq
This looks like an $s$-channel propagator for a particle at rest of mass 
$[2M + (-\mu+1/a)/Mr_0]$, and in fact, for $\mu=0$, corresponds exactly to the 
dibaryon proposed in \cite{DBK} to reproduce the scattering 
due to a short range potential.  We see that using the dibaryon is as good as (but no better than)
 carrying out the effective field theory calculation to $O(p^1)$. The subleading 
corrections can be accounted for by including the subleading part of the $ C_4(\mu) 
p^4$ vertex proportional to $r_1$, and which occurs at  $O(p^2)$.   This dibaryon  
was recently used with great success in the three-body problem \cite{Bvk}.

\subsection{Luke-Manohar velocity scaling}
\label{sec:2d}

In ref. \cite{LMa} a rescaling of fields was shown to eliminate the
heavy mass scale $M$ from the Lagrangian, and allow a simple power counting in
terms of the velocity $v$ of the interacting nucleons.  One finds that a four fermion
operator scales as $v$, and is therefore irrelevant as $v\to 0$.  
It is instructive to see
why this conclusion is avoided in the theory with large scattering length, 
where the interaction
has a large effect on scattering at the low momentum $p\sim 1/|a|$.

The rescaled theory is
described in terms of dimensionless energy and momentum variables, ${\cal
E}=E/Mv^2$ and ${\cal P}=p/Mv$, corresponding to a rescaling of the 
spacetime coordinates 
$T= t M v^2$, $X=x M v$.  The nucleon fields are  rescaled as $\Psi=\tilde N/(M v)^{3/2}$, 
in order to give a
canonical energy term.  Performing the Luke-Manohar scaling on our toy theory
in a naive manner yields the action
\beq
S=\int d^3 X dT\ \ 
{\Psi^\dagger} \( i\partial_T + {1\over 2}\nabla^2_X\) \Psi
+C_0 M^2 v ({\Psi^\dagger}\Psi )^2
 +...
\eqn{yukeffLM}
\eeq
This demonstrates the familiar result that a contact interaction
($\delta$-function potential)  is an irrelevant operator in nonrelativistic
scattering in $D=4$ dimensions, since the interaction vanishes as $v\to 0$.
This is certainly true when the underlying theory is perturbative, but it fails
when it is strongly coupled and there is a large scattering length.  In that
case we should be using the running coupling constants discussed in the
previous section, keeping in mind that the renormalization scale $\mu$ must
also be rescaled as $\hat\mu=\mu/Mv$,  since it is treated on the same footing as 
 momenta.
Replacing $C_0$ in the above expression with the value of the running coupling
$C_0(\mu)$ from \eq{c0run} we find the rescaled interaction
\beq
\int d^3 X dT \ 
{4\pi M v  \over -\hat \mu M v + 1/a}({\Psi^\dagger} \Psi)^2
 +...
\eqn{yukeffLMresc}
\eeq
where $\hat\mu\sim 1$.  Now it is evident once again that the limit $a\to\infty$
corresponds to tuning $C_0$ to a UV fixed point.  Near that fixed point, the
four fermion interaction is a {\it relevant} operator (it doesn't decrease as
$v$ gets smaller), which is why it has a large effect on scattering at low velocity. 
Only for the very low velocities $v\ltap 1/Ma$  does it become irrelevant, at which point
perturbation theory is once again justified.

\section{The effective field theory expansion for realistic NN scattering}
\label{sec:3}

We now turn to the problem of interest, realistic low momentum  nucleon-nucleon
scattering.  The goal of an effective field theory treatment is to provide a
systematic expansion for computing $NN$ phase shifts at low momenta, which
means that the phenomenology of the two nucleon system  can be approximated to
any desired accuracy.   While an effective field theory treatment will never in
practice compete with the extremely precise phenomenological potential model
descriptions of the $NN$ phase shifts, it will provide a framework to
compute  relativistic effects and inelastic processes, as well as to reliably
estimate errors at any order in the expansion.  Furthermore, as it is a perturbative
expansion, the effective field theory approach will hopefully be useful for
many-body systems for which potential models are computationally intractable.

Realistic $NN$ scattering is more complicated than our toy model for several
reasons: the inclusion of explicit pion fields,  the spin and
isospin degrees of freedom, and the need to include relativistic corrections in
the power counting scheme.  Nevertheless, we will show that the momentum expansion
for this theory closely follows the discussion in \S\ref{sec:2b}, and that for
$p\ltap m_\pi$ the expansion improves as the pion mass gets smaller.  This
is an important result, as it means that there is a region of QCD parameters
for which our results hold with high accuracy. (We do not, however, take the
pion mass so small that $am_{\pi}$ is also small.)

Since we are performing a momentum expansion, it is important to understand
the scale at which it fails.  In the 1-nucleon sector, the expansion is in powers of
$(m_\pi/\Lambda_\chi)$
and $(p/\Lambda_\chi)$, where $\Lambda_\chi = 4\pi f=1.6 {\rm GeV}$, ($f=132\ \MeV$ being 
the pion decay constant).  However, we find that the chiral
expansion for $NN$ scattering entails the new scale,
\beq
\Lambda_{NN} \equiv {8\pi f^2\over g_A^2 M } = 300\,\MeV\ .
\eqn{lnndef}
\eeq
It is disturbing that
$\eta_\pi\equiv (m_\pi/\Lambda_{NN})= 0.46$, suggesting that our expansion
will not converge rapidly.  However, if one looks at the Yukawa piece of
the one pion exchange (OPE) potential in the $\si$ channel, it only binds nucleons for
$\eta_\pi  \ge 1.7$,  while for  the true value $\eta_\pi = .46$, the $NN$ phase
shift from the OPE Yukawa potential never gets larger than $\sim .25$ radians. Thus
there is empirical evidence that the chiral expansion parameter for $NN$ scattering
is about $30\%$, and it is reasonable that for momenta $p\sim m_\pi$ the expansion
will converge fairly quickly.

\subsection{The chiral expansion}
\label{sec:3a}

We need to consider a theory of nucleons interacting with pions and with
themselves through contact interactions consistent with chiral symmetry.
Nucleons are described by isodoublet fields $N$, while the pions are described
by the field
\beq
\xi(x) = e^{i\Pi/f}\ ,\qquad \Pi = \(\matrix{{\pi^0\over\sqrt{2}} & \pi^+\cr
\pi^- &-{\pi^0\over\sqrt{2}} \cr}\)
\eeq
and $f$ is the pion decay constant normalized to be
\beq
f=132\ \MeV\ .
\eqn{fval}
\eeq
Under $SU(2)_L\times SU(2)_R$ chiral symmetry the fields transform as
\beq
\xi \to L \xi U^\dagger =  U  \xi R^\dagger\ , \qquad
\Sigma\equiv \xi^2\to L\Sigma R^\dagger\ ,\qquad
N\to U N\  ,
\eqn{pitrans}
\eeq
where $L$, $R$ are constant $SU(2)$ matrices and $U(x)$ is a
pion-dependent $SU(2)$ matrix.  
The vector and axial-vector pion currents are given by
\beq
V_\mu = {1\over 2}(\xi \partial_\mu\xi^\dagger + \xi^\dagger\partial_\mu\xi)\
,\qquad
A_\mu = {i\over 2}(\xi \partial_\mu\xi^\dagger - \xi^\dagger\partial_\mu\xi)
\eeq
where the axial current $A_\mu$ and the chiral covariant derivative
$D_\mu=(\partial_\mu+V_\mu)$ transform linearly as
\beq
A_\mu \to U  A_\mu U^\dagger \ ,\qquad D_\mu \to U  D_\mu
U^\dagger \ .
\eeq

In the 1-nucleon sector, the chiral Lagrangian is then given by
\beq
\CL =  N^\dagger (iD_0+ \vec D^2/2M) N + {f^2\over 8} \Tr \partial_\mu
\Sigma^\dagger  \partial^\mu \Sigma +
   {f^2\over 4}\omega\Tr M_q (\Sigma + \Sigma^\dagger)   + g_A N^\dagger \vec
A\cdot\vec\sigma N +\dots
\eqn{chilag}
\eeq
where $M_q$ is the quark mass matrix diag$(m_u,m_d)$, $\omega$ has dimensions
of mass with $m_\pi^2=\omega(m_u+m_d)$, and the ellipses refers to all
additional operators.

When considering the 2-nucleon sector, one must include local four nucleon
operators in the effective lagrangian.
Due to spin and isospin degrees of freedom, the number of four fermion
operators grows rapidly with the number of derivatives.  To the order
we will be working, we need only consider operators without pion fields
and with $\le 2$ derivatives, or no derivatives and one insertion of
the quark mass matrix. Rather than writing
down the operators in the chiral Lagrangian, it is simplest to identify their
matrix elements between particular partial waves.  
With no derivatives, these
contact interactions  only effect  $S$-waves.  Thus there are
two independent $C_0$ operators associated with the $\si$ and $\siii$ channels
\cite{Weinberg1}.  Denoting the partial waves as
\beq
\ket{\alpha} = \ket{SLJm,E}
\eeq
we normalize the two $C_0^{(\gamma)}$ operators by defining the Born amplitude
for $\CL_{C_0}$, the four fermion interactions involving no derivatives, as
\beq
 \bra{\alpha}  \CL_{C_0}\ket{\beta} = -\left( {M p \over 2 \pi} \right)
\sum_\gamma
C_0^{(\gamma)}\delta_{\alpha\gamma}\delta_{\beta\gamma} \ ,\qquad
\gamma=\{\si,\siii\}\ .
\eqn{czeroops}
\eeq

There are seven independent two derivative contact interactions \cite{KoMany}.
These operators change orbital angular momentum by
 $\Delta L=0$ or $\Delta L=2$.
Therefore, in analogy to \eq{czeroops}, we can define seven $C_2^{(\gamma)}$
couplings by
\beq
 \bra{\alpha}  \CL_{C_2}\ket{\beta} &=&  -\left( {M p \over 2 \pi} \right)
p^2 \[\sum_{\gamma}
C_2^{(\gamma)}\delta_{\alpha\gamma}\delta_{\beta\gamma}  +
(C_2^{(\siii-\diii)}\delta_{\alpha,\siii}\delta_{\beta,\diii}
+ \alpha\leftrightarrow\beta)\]\ , \nonumber\\
\gamma&=&\{{}^1\kern-.14em S_0,  {}^1\kern-.14em P_1,  {}^3\kern-.14em S_1,
{}^3\kern-.14em P_0,  {}^3\kern-.14em P_1, {}^3\kern-.14em P_2\}\ .
\eqn{ctwoops}
\eeq
As before, $p^2\equiv ME$.

At the same order as the $C_2$ operators are four fermion operators with a
single insertion of the quark mass matrix $M_q$. If we ignore isospin
violation, then $M_q\propto m_\pi^2$ times a unit matrix in isospin space, 
and so these operators have the same
structure as the $C_0^{(\gamma)}$ operators.  We parametrize them as
\beq
 \bra{\alpha}  \CL_{D_2}\ket{\beta} = -\left( {M p \over 2 \pi} \right) m_\pi^2\sum_\gamma
D_2^{(\gamma)}\delta_{\alpha\gamma}\delta_{\beta\gamma} \ ,\qquad
\gamma=\{\si,\siii\}\ .
\eqn{dtwoops}
\eeq

Since the  $\si$ and $\siii$ scattering lengths are large compared with $1/m_{\pi}$,
the power counting in these channels is the same as for the example of \S2:
for  $\mu\sim m_\pi$,
$C_0^{(\gamma)}(\mu)\propto 1/\mu$,  $C_2^{(\gamma)}(\mu)\propto 
1/\mu^2$ and $D_2^{(\gamma)}(\mu)\propto 
1/\mu^2$ for $\gamma=\{\si,\siii\}$.   A single  exchange of a potential pion
contributes to the amplitude a factor of
\beq
i{g_A^2\over 2 f^2} {\bfq\cdot\sigma^1 \bfq\cdot\sigma^2\over \bfq^2+m_\pi^2}\tau^1\cdot\tau^2
\eeq
which scales as $O(1)$ in the chiral expansion.  Thus one pion exchange occurs at the same
order as an insertion of $C_2p^2$ or $D_2 m_{\pi}^2$ in the  $\si$ and $\siii$ channels, and pion exchange 
may be treated perturbatively.  A new feature of the theory with pions is that this scaling 
behavior breaks down not only at low momentum, $p\sim 1/|a|$, but also at sufficiently
high momentum.  This can be seen by examining the RG equations for $C_0^{(\gamma)}$.
\begin{figure}[t]
\centerline{\epsfysize=1 in \epsfbox{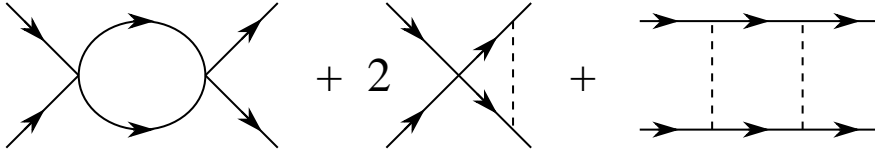}}
\noindent
\caption{\it Leading contributions to the $\beta$-functions for $C_0^\gamma$,
including pions}
\label{betapi}
\vskip .2in
\end{figure}
The exact beta function for the $C_0^{(\gamma)}$ coefficients comes from the graphs of
Fig.~\ref{betapi}, and one finds
\beq
\beta_0^{(\si)} =\mu {{\rm d}C_0^{(\si)}\over {\rm d}\mu\ \ } =  {M\mu\over 4\pi}
\left\{\(C_0^{(\si)}\)^2+ 2 C_0^{(\si)}{g_A^2\over 2 f^2}+ \({g_A^2\over 2
f^2}\)^2\right\} ,
\eqn{beta010}
\eeq
\beq
\beta_0^{(\siii)} =\mu {{\rm d}C_0^{(\siii)}\over {\rm d}\mu\ \ } =  {M\mu\over
4\pi} \left\{\(C_0^{(\siii)}\)^2+  2 C_0^{(\siii)}{g_A^2\over 2 f^2}+ 9 
\({g_A^2\over 2 f^2}\)^2\right\} .
\eqn{beta030}
\eeq

Solving the RG equation for the singlet channel \eq{beta010} with the boundary
condition $C_0^{(\si)}(0)=4\pi a_1/M$, where $a_1$ is the $\si$
scattering length,  we find
\beq
C_0^{(\si)}(\mu) =  -{4\pi\over M\mu}\({1\over 1-{1\over \mu(a_1 + 1/\Lambda_{NN})}} 
+ {\mu\over \Lambda_{NN}}   \) ,
\eeq
where $\Lambda_{NN}$ is given in \eq{lnndef}.
  Since $\Lambda_{NN}\gg 1/|a_1|$, for 
$\mu\gg 1/|a_1|$ we have
\beq
C_0^{(\si)}(\mu)\simeq -{4\pi\over M\mu}\(1+{\mu\over
\Lambda_{NN}}\)\ ,
\eeq
and so the power counting changes for $\mu\gtap \Lambda_{NN}$;  in fact the 
UV fixed point toward which $C_0^{(\si)}$ is driven largely cancels the $\delta$-function component
of the single pion exchange in the $\si$ channel.  Similarly, power counting 
is found to change at $\mu\sim\Lambda_{NN}$  in the $\siii$ 
channel as well. As a 
result, the power counting developed in \S2, with one pion exchange treated as $O(1)$,  
works only up 
to $p\sim \Lambda_{NN}$.  At that point one pion exchange and a $C_0$ insertion become equally 
important.   The power counting that we follow in this paper is therefore  expected 
to fail completely at momenta on the order of $\Lambda_{NN}$.

Up to now we have only discussed the scaling of the $C_2$ interactions in the $\si$ and $\siii$ 
channels.  The $1/\mu^2$ scaling 
of these couplings is due to multiplicative renormalization 
by $C_0$.  Such renormalization does not occur for  $C_2$  in any of the $P$-wave channels, and so 
these couplings will be $O(p^0)$.  Thus the leading contribution to the $P$-wave (and all higher partial 
wave) amplitudes occurs at $O(1)$ and is simply  a single pion exchange (the Born approximation OPE).  
It is well-known that 
this accounts very well for the observed phase shifts.  The subleading contribution to the $P$-wave 
amplitude is due to the exchange of two potential pions at $O(p^1)$, while the $C_2p^2$ operators enter
at $O(p^2)$, along with the two-loop graph with the exchange of three pions.

The transition operator coefficient, 
$C_2^{(\siii-\diii)}$ is different from the others as it is multiplicatively 
renormalized by $C_0^{(\siii)}$, but its beta function is half as big as that for $C_2^{(\siii)}$.  
For small $\mu$, $C_2^{(\siii-\diii)}(\mu) \sim 1/\mu $, and it enters the calculation at $O(p^1)$. An important aspect of the $O(p^1)$ calculation of the
$\si$ and $\siii$ scattering amplitudes is the absence of  new undetermined
coefficients $C_4^{(\si)}$ and $C_4^{(\siii)}$. The renormalization group
scaling determines the leading behavior of these couplings. For small $\mu$
(but $|a|\mu \gg 1$), $C_4^{(\gamma)}(\mu)=-(M\mu / 4 \pi)C_2^{(\gamma)}(\mu)^2$.

\begin{figure}[t]
\centerline{\epsfysize=1 in \epsfbox{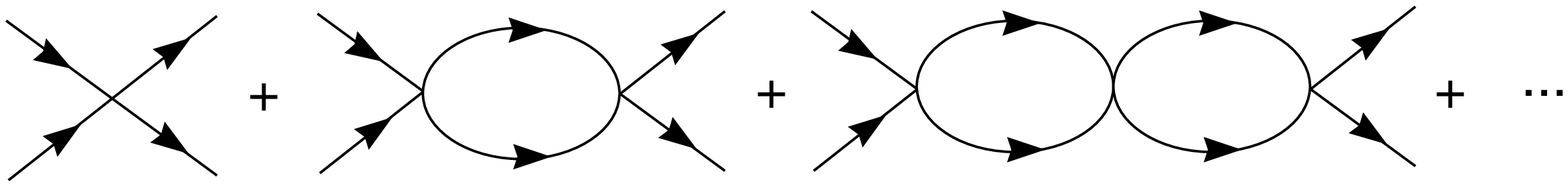}}
\noindent
\caption{\it Graphs contributing to the leading amplitude $\CA_{-1}$.}
\label{bubbleminus}
\vskip .2in
\end{figure}

\begin{figure}[t]
\centerline{\epsfysize=3 in \epsfbox{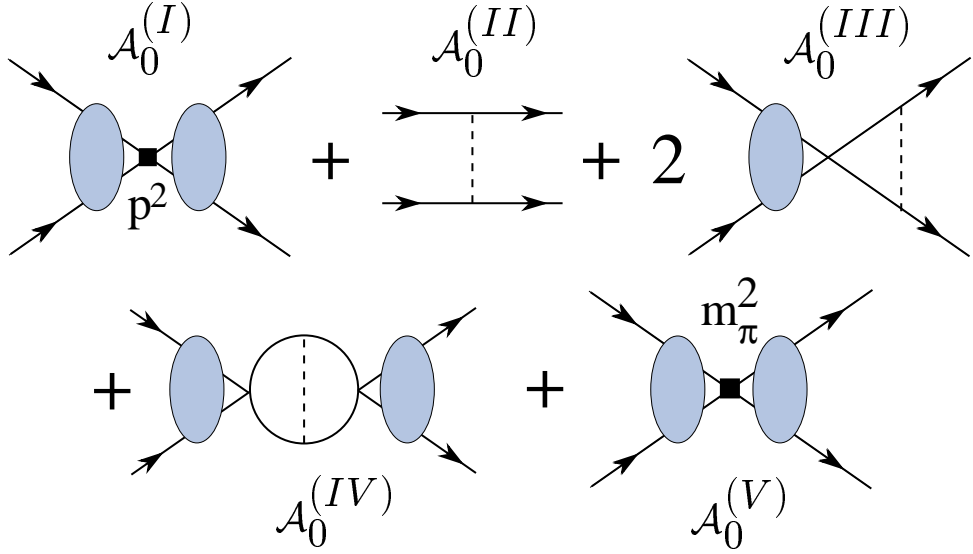} }
\noindent
\caption{\it Graphs contributing to the subleading amplitude $\CA_{0}$; 
the shaded ovals are defined in Fig.~\ref{FG1S0_m1}. }
\vskip .2in
\label{qzero}
\end{figure}

It is now possible to perform a systematic chiral expansion
for $NN$ scattering amplitudes in the situation where $m_\pi$ is small, 
(but $|a| m_\pi$ is not).  
Denoting our expansion parameter by $Q$,
representing equally $p$ or $m_\pi$, and taking $\mu \sim Q$ the leading amplitude is $O(Q^{-1})$ and
is simply the bubble sum of $C_0^{(\gamma)}$ interactions, shown in Fig.~\ref{bubbleminus}.  
The subleading contribution is $O(Q^0)$, and involves
an insertion of either a single pion exchange, a $C_2$ operator, or a $D_2$
operator, dressed to all orders by the $C_0$ interaction, as shown in 
Fig.~\ref{qzero}.
At $O(Q^1)$ one finds both  two pion exchange (involving both the two pion
vertex as well as iterated one OPE), four derivative operators, and so forth.
It may be tempting to sum up the OPE potential exactly as is typically done
(e.g., ref. \cite{Weinberg1}), but it is not consistent to do so
while leaving out the arbitrarily higher derivative contact interactions that
are required to renormalize the pion ladders.  By neglecting these higher derivative
terms one is making a model for the short distance physics, and the result
one gets is not any more accurate for having included the multiple pion exchange.  
%

\subsection{The $\si$ channel to subleading order}
\label{sec:3c}

\begin{figure}[t]
\centerline{\epsfysize=3 in \epsfbox{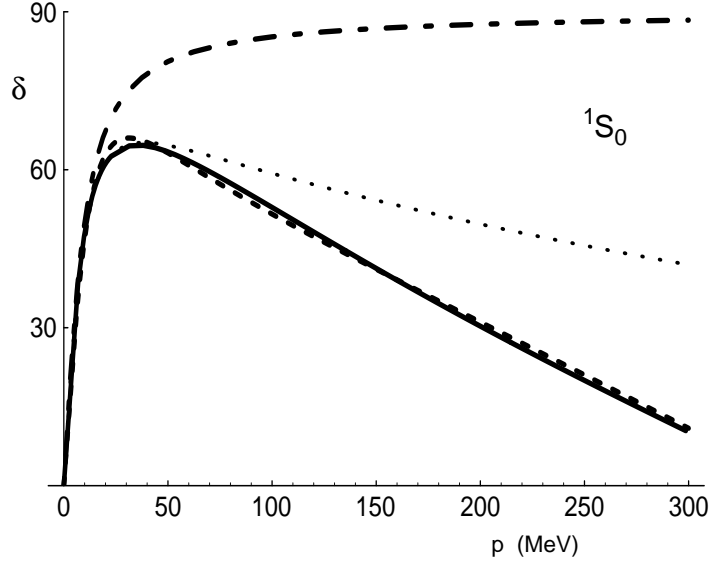}}
\noindent
\caption{\it The phase shift $\delta$ for the $\si$ channel.
The dot-dashed curve is the  one parameter fit at $O(Q^{-1})$,
that reproduces the scattering length.
The dotted  and dashed curves are the  fits 
at $O(Q^0)$ in \eqsii{dzfit}{numfit} respectively.
The dashed curve corresponds to fitting $\delta$ 
between $0 < p < 200\ {\rm MeV}$, while the 
dotted curve corresponds to fitting the 
scattering length and effective range.
The solid line shows the results of the 
Nijmegen partial wave  analysis.
}
\label{singfit}
\vskip .2in
\end{figure}

In the isotriplet $\si$  channel the nucleons are exclusively in an
$S=L=J=0$ state.
The leading $O(Q^{-1})$ contribution  to scattering 
in this channel is from the bubble chain 
of $C^{(\si)}_0$ operators, as shown in Fig.~\ref{bubbleminus}
\beq
\CA_{-1}=
-{ C^{(\si)}_0\over  1 + C^{(\si)}_0 {M\over 4\pi}  \left( \mu + i p \right) 
}
\ .
\eqn{simi}
\eeq
Since we are computing the amplitude between plane wave states, we omit from $\CA$
the normalization factor $(Mp/2\pi)$ appearing in \eqs{czeroops}{dtwoops}.
There is only one unknown parameter that needs to be fit to data
and for simplicity
we will require that the amplitude at this order 
reproduce the $\si$ ($np$) scattering length  
$a_1 = -23.714\pm 0.013\ {\rm fm}$. 
This determines $C_0^{(\si)}(m_{\pi})=-3.51\fm^2$,
where we have chosen to renormalize at
$\mu=m_\pi$ and the behavior 
of the phase shift over a range of momenta resulting from this fit
is shown in Fig.~\ref{singfit}.

At next order $Q^0$ there are contributions from insertions of 
higher dimension
local operators and also from potential pion exchange. The 
 $O(Q^0)$  contribution to the amplitude is
written as a sum of the five terms, ${\cal A}_0={\cal A}_0^{(I)}+
{\cal A}_0^{(II)}+{\cal A}_0^{(II)}+{\cal A}_0^{(IV)}+{\cal A}_0^{(V)}$.
The  local operators at this order involve either two spatial derivatives, 
$C_2^{(\si)}$,
or one insertion of the light quark mass matrix, $D_2^{(\si)}$.
Expressions for the graphs shown in Fig.~\ref{qzero}
which 
we presented in ref. \cite{KSWb}
are
\beq
 {\cal A}_0^{(I)} &=& 
-C_2^{(\si)} p^2
\left[ {\CA_{-1}\over C_0^{(\si)}  } \right]^2
\ \ \ ,
\nonumber\\
 {\cal A}_0^{(II)} &=&  \left({g_A^2\over 2f^2}\right) \left(-1 + {m_\pi^2\over
4p^2} \ln \left( 1 + {4p^2\over m_\pi^2}\right)\right)
\ ,
\nonumber\\
 {\cal A}_{0}^{(III)} &=& {g_A^2\over f^2} \left( {m_\pi M{\cal A}_{-1}\over 4\pi}
\right) \Bigg( - {(\mu + ip)\over m_\pi}
+ {m_\pi\over 2p} \left[\tan^{-1} \left({2p\over m_\pi}\right) + {i\over 2} \ln
\left(1+ {4p^2\over m_\pi^2} \right)\right]\Bigg)
\ ,
\nonumber\\
{\cal A}_0^{(IV)} &=& {g_A^2\over 2f^2} \left({m_\pi M{\cal A}_{-1}\over
4\pi}\right)^2 \Bigg(-\left({\mu + ip\over m_\pi}\right)^2
+ \left[ i\tan^{-1} \left({2p\over m_\pi}\right) - {1\over 2} \ln
\left({m_\pi^2 + 4p^2\over\mu^2}\right) + 1\right]\Bigg)
\ ,
\nonumber\\
{\cal A}_0^{(V)} &=& - D^{(\si)}_2 m_\pi^2 
\left[ {\CA_{-1}\over C_0^{(\si)}  }\right]^2
\ .
\end{eqnarray}

The two-loop diagram in $\CA_0^{(IV)}$ involving the exchange of a potential pion
between two contact terms is divergent in 
both three and four dimensions.
In the {\it PDS} scheme we subtract the poles in three and 
four dimensions leaving this graph logarithmically as
well as power-law 
dependent on the renormalization point $\mu$.
As the coefficient of the four-dimensional divergence is proportional 
to the mass of the pion squared, the required isospin conserving counterterm 
with coefficient $D^{(\si)}_2 (\mu)$
depends on the sum of the light quark masses, $m_u+m_d$  
and gives rise to ${\cal A}_0^{(V)}$.
In addition, for convenience
we have 
absorbed some of the $\mu$-independent terms from ${\cal A}_0^{(IV)}$
into the definition of $D^{(\si)}_2 (\mu)$.
At this order there are three unknown counterterms
that need to be fit to data, $C_0^{(\si)}(\mu), C_2^{(\si)}(\mu)$ and 
$D^{(\si)}_2 (\mu)$.
As the amplitude can be written as a function of 
$C_0^{(\si)}(\mu) + m_\pi^2  D^{(\si)}_2 (\mu)$, the dependence of observables upon
$C_0^{(\si)}(\mu)$ and $ D^{(\si)}_2 (\mu)$ individually is an artifact of the 
perturbative expansion, and is indicative of the size of higher order effects.
Conventionally, the scattering data in the $NN$ sector is  
presented in terms of phase shifts.
In this channel, the phase shift is simply related to the  
amplitude $\CA$ by
\beq
S= e^{2i\delta} = 1+ i{M p\over 2\pi}\CA
\ \ \ ,
\eeq
leading to
\begin{eqnarray}
\delta & = & 
{1\over 2i} \ln\left( 1 + i {Mp\over 2\pi}\CA \right)
\ .
\end{eqnarray}
Expanding both sides to a given order in $Q$ with 
$\delta = \left( \delta^{(0)} + \delta^{(1)} + \ldots
\right) $ 
gives
\begin{eqnarray}
\delta^{(0)} & = & 
{1\over 2i} \ln\left( 1 + i {Mp\over 2\pi}\CA_{-1} \right)
\ ,\qquad
\delta^{(1)} = {M p \over 4 \pi} \left( {\CA_0\over 1 + i {Mp\over 2\pi}\CA_{-1} }
\right)
\ .
\end{eqnarray}
Here superscripts denote the order in Q. 
Our expression for the amplitude $\CA$ gives an $S$-matrix that 
is unitary up to the order we have computed, i.e.
if the amplitude is computed up to $O(Q^k)$, then 
$S^\dagger S = 1+O(Q^{k+2})$.  

It is convenient to choose $\mu = m_{\pi} $. 
Expressions for the scattering length and
effective range are determined from the expansion, 
$p \cot \delta = ip +4\pi /M{\cal A}_{-1}- 4\pi {\cal A}_0 /M{\cal A}_{-1}^2$, 
which yields to the order we
are working,
\begin{equation}
{1 \over a_1} = \left( m_\pi+{4\pi \over MC_0^{(\si)}(m_{\pi})}\right)-
{4 \pi D_2^{(\si)}(m_{\pi})m_{\pi}^2 \over M(C_0^{(\si)}(m_{\pi}))^2} 
\ \ \  ,
\end{equation}
and
\begin{equation}
r_0={8\pi C_2^{(\si)}(m_{\pi}) \over M(C_0^{(\si)}(m_{\pi}))^2}+{g_A^2M \over 12\pi f^2}\left(
1+{16\pi \over C_0^{(\si)}(m_{\pi})m_{\pi}M}+{96 {\pi}^2 \over (C_0^{(\si)}(m_{\pi})m_\pi M)^2} \right) .
\end{equation}
The choice
\beq
C_0(m_{\pi})=-3.63~{\rm fm}^2\ ,\qquad D_2(m_{\pi})\equiv 0\ ,\qquad
C_2(m_{\pi})=2.92~{\rm fm}^4\ ,
\eqn{dzfit}\eeq
 is consistent
with the experimental values of the scattering length and effective range 
\footnote{At this order there is  ambiguity in these values since to the order we are 
working only the linear combination, $C_0(m_{\pi})+m_{\pi}^2D_2(m_{\pi})$, is determined. 
At higher order the $C_0$ and $D_2$ operators are distinguished by a contribution to an 
$N^\dagger N N^\dagger N\pi\pi$ vertex proportional to $D_2$.}. 
About $43\%$ of the effective range is due to $C_2(m_{\pi})$. The phase shift resulting
from these parameters is shown in Fig.~\ref{singfit}.

Alternatively, we consider fitting
the phase shift from the 
Nijmegen partial wave  analysis \cite{nijmegen} 
over the momentum range $p\le 200 \MeV$, treating  $C_0^{(\si)}$,
 $D_2^{(\si)}$ and  $C_2^{(\si)}$ as free parameters.  
The results are 
\begin{equation}
C_0^{(\si)}(m_{\pi})=-3.34\fm^2
\ ,\qquad
D_2^{(\si)}(m_{\pi})=-0.42\fm^4
\ ,\qquad
C_2^{(\si)}(m_{\pi})=3.24\fm^4
\ \ \  ,
\eqn{numfit}
\end{equation}
which give the phase shift plotted in  Fig.~\ref{singfit}. 
As is apparent from  Fig.~\ref{singfit}, 
the agreement of the phase shift with data 
is excellent at quite large values of $p$. Furthermore,
the coupling $C_0^{(\si)}(m_{\pi})$
is close to its leading order value (in the limit of large scattering length),
$-(4\pi/M m_\pi)=-3.7\fm^2$, and $C_2^{(\si)}(m_{\pi})$ is also near its  expected size, suggesting that
our expansion is valid in this channel. However, for
$p>100\MeV$ the magnitude of the ratio
${\cal A}_0/{\cal A}_{-1}$ is greater than $\sim 0.5$ and it is
difficult to justify the approximations we have made, e.g.
neglecting terms suppressed by  $({\cal A}_0/{\cal A}_{-1})^2$. 
The difference between the two  fitting procedures 
is attributable to effects higher order in our expansion.
%

\subsection{The ${}^3S_1-{}^3D_1$  channels to subleading order}
\label{sec:3e}

The analysis of the isosinglet $\siii-\diii$ channel is   richer than the 
$\si$ channel since there 
are two different orbital angular momentum states  involved.
The power counting in the $\siii$ channel is the same as for the 
$\si$ channel.
However, we must also consider how the coefficients of the operators contributing to 
scattering in the $\diii$ channel
and the coefficients of the operators that  give rise to  
$\siii-\diii$ mixing behave under
renormalization group scaling and at what order in the $Q$ expansion they
contribute to observables.
Firstly,  operators  between two $\diii$ states are not renormalized by the 
leading operators, which project out only $\siii$ states.
Further, they involve a total of four spatial derivatives, two on the incoming 
nucleons, and  two on the out-going nucleons.   
Therefore, such operators contribute
at $O(Q^4)$, and can be safely neglected.
Consequently, amplitudes for scattering from an $\diii$ state into an $\diii$ state
are dominated by single potential pion exchange which contributes at $O(Q^0)$.
Single pion exchange describes well this partial 
wave in the  momentum range we are considering.
Secondly, operators connecting $\diii$ and $\siii$ states involve 2 spatial 
derivatives (acting on the $\diii$ state) 
and are renormalized by the leading  operators, but only on the 
$L=0$ ``side'' of the operator.
Therefore the coefficient of this operator, $C_2^{(\siii-\diii)}\sim 1/\mu$,
and contributes at $O(Q^1)$. 
Hence it can also be neglected
at the order we are working.
Thus,  mixing between $\diii$ and $\siii$ states is dominated by
single potential pion exchange dressed by a bubble chain of $C_0^{(\siii)}$ operators
and a parameter free prediction for this mixing as a function of momentum
exists at $O(Q^0)$.

We denote the amplitude at $O(Q^n)$ by $\CA_{n[LL^\prime]}$, where $L$ and $L^\prime$ are the 
initial and final orbital angular momenta.  
As in the $\si$ channel,  we omit from $\CA$
the normalization factor $(Mp/2\pi)$ appearing in \eqs{czeroops}{dtwoops}
since we are computing the amplitude between plane wave states.
At leading $O(Q^{-1})$ in the expansion there is a contribution only to the 
$\siii$ partial wave:
\begin{eqnarray}
\Czzm  = 
-{ C^{(\siii)}_0\over  1 + C^{(\siii)}_0 {M\over 4\pi}  \left( \mu + i p \right) 
}\ ,\qquad
\Cztm\  = \  \Ctzm\  =\  \Cttm \ = \ 0
\  .
\eqn{simib}
\end{eqnarray}
At $O(Q^0)$ there are contributions from  graphs of the same form as in 
the amplitude for $\si$ scattering, shown in Fig.~\ref{qzero}.
Using the same identification of graphs  as in the $\si$ channel, and the similar notation, ${\cal A}_{0[L,L^{\prime}]}={\cal A}_{0[L,L^{\prime}]}^{(I)}
+...$,we find that 
\begin{eqnarray}
\Czzz^{(I)}  =  
-C_2^{(\siii)} p^2
\left[ {\Czzm\over C_0^{(\siii)}  } \right]^2
\ ,\qquad
\Cztz^{(I)} \  = \  \Ctzz^{(I)} \ =\  \Cttz^{(I)} 
\ =\ 0
\  ,
\end{eqnarray}
arising from a single insertion of the local operator 
involving two spatial derivatives. 
Single potential pion exchange in the $t$-channel gives
\begin{eqnarray}
\Czzz^{(II)} \ & = &\  
-{g_A^2\over 2 f^2} \left[ 1 - {m_\pi^2\over 2 p^2} Q_0(z)\right]
\ \ \ ,
\nonumber\\
\Cztz^{(II)} \ & = &\  \Ctzz^{(II)} \ =\ 
-{g_A^2  \over \sqrt{2}  f^2} 
\left[ Q_0(z) + Q_2(z)-2 Q_1(z) \right]
\ \ \ ,
\nonumber\\
\Cttz^{(II)} \ & = &\  
-{g_A^2  m_\pi^2\over 4 f^2 p^2} 
\left[ Q_2(z) + {6 p^2\over 5m_\pi^2}\left(Q_1(z)-Q_3(z)\right)\right]
\ \ \ ,
\end{eqnarray}
where $z = 1+m_\pi^2/(2 p^2)$ and  $Q_k (z)$ denotes
the $k$-th order irregular Legendre function,  
\begin{eqnarray}
Q_0 = {1\over 2}\log\left({z+1\over z-1}\right)\ ,\quad
Q_1   =    z Q_0-1\ ,
\quad
(n+1) Q_{n+1}  =  (2 n+1) z Q_n - n Q_{n-1}\ .
\end{eqnarray}
The contribution from single pion exchange 
across the end of a bubble chain of operators
with coefficient
$C_0^{(\siii)}$ is
\begin{eqnarray}
\Czzz^{(III)} &=& {g_A^2\over f^2} \left( {m_\pi M \Czzm\over 4\pi}
\right)\ 
\Bigg( - {(\mu + ip)\over m_\pi}
+ {m_\pi\over 2p} \left[\tan^{-1} \left({2p\over m_\pi}\right) + {i\over 2} \ln
\left(1+ {4p^2\over m_\pi^2} \right)\right]\Bigg)
\ \ \ ,
\nonumber\\
\Cztz^{(III)} \ & = & \ \Ctzz^{(III)} 
\nonumber\\
& = & 
{g_A^2\over \sqrt{2} f^2}
\left({ M  \Czzm\over 4 \pi}\right) p^2 
\left[ -{3 m_\pi^3\over 4 p^4} 
+ {m_\pi^2\over 8 p^5} (3 m_\pi^2 + 4 p^2)
\tan^{-1}\left({2 p\over m_\pi}\right)
\right.
\nonumber\\
& & \left.
\qquad\qquad + i \left( -{3 m_\pi^2\over 4 p^3} 
+ {1\over 2 p} 
+ {m_\pi^2\over 4 p^3} 
\left( 1+ {3 m_\pi^2\over 4 p^2} \right)
\log\left(1+{4 p^2\over m_\pi^2}\right)
\right)
\right]
\ \ \ ,
\nonumber\\
\Cttz^{(III)} \ & =& \  0
\ ,
\end{eqnarray}
while
\begin{eqnarray}
\Czzz^{(IV)} &=& 
{g_A^2\over 2f^2} 
\left({m_\pi M \Czzm\over 4\pi}\right)^2 
 \Bigg(-\left({\mu + ip\over m_\pi}\right)^2
+ i\tan^{-1} \left({2p\over m_\pi}\right) - {1\over 2} \ln
\left({m_\pi^2 + 4p^2\over\mu^2}\right) + 1\Bigg)\ ,
\nonumber\\
\Cztz^{(IV)} \ &=& \ \Ctzz^{(IV)} \ =\ 
\Cttz^{(IV)} \ = \  0
\ ,
\end{eqnarray}
is from pion exchange  between two chains of operators with coefficients
$C_0^{(\siii)}$.
Finally, a single insertion of the quark mass matrix 
leads to 
\begin{eqnarray}
\Czzz^{(V)} = - D^{(\siii)}_2 m_\pi^2 
\left[ {\Czzm\over C_0^{(\siii)}  }\right]^2\ ,\qquad
\Cztz^{(V)} \ = \ \Ctzz^{(V)} \ =\ 
\Cttz^{(V)} \ = \  0
\  .
\end{eqnarray}
Again part of the subtraction point independent contribution to $\Czzz^{(IV)}$
has been absorbed into $\Czzz^{(V)}$.

The S-matrix
in this channel is usually expressed in terms of two phase shifts,
$\delta_{0}$ and 
$\delta_{2}$, and a mixing angle $\varepsilon_1$,
\begin{equation}
S =  1\ +\ i{pM \over 2\pi}{\cal A}=\left( \begin{array}{ll}
e^{2i\delta_{0}} \cos 2\varepsilon_{1}  & ie^{i(\delta_{0} + \delta_{2})}
\sin 2\varepsilon_1\\
ie^{i(\delta_{0} + \delta_{2})} \sin 2\varepsilon_1 & e^{2i\delta_{2}} \cos
2\varepsilon_1 \end{array} \right) 
\  ,
\end{equation}
and like the $\si$ channel we will expand $S$ order by order
in $Q$.

As $\siii-\diii$ mixing has vanishing contribution at order 
$O(Q^{-1})$ the mixing parameter
$\varepsilon_{1}$ starts at $O(Q^1)$, the same holds true for $\delta_2$
(the explicit factor of $p$ in the relation between $S$ and $\CA$ 
increases the order by $1$).
Writing each of the parameters as an expansion in $Q$,
\begin{eqnarray}
\delta_0 & = &  \delta_0^{(0)}\ +\ \delta_0^{(1)}\ +\ ...
\ ,\qquad
\delta_2 =  \delta_2^{(0)}\ +\ \delta_2^{(1)}\ +\ ...
  \ ,\qquad 
\varepsilon_{1} =  \varepsilon_{1}^{(0)}\ +\ \varepsilon_{1}^{(1)}\ +\ ...
\ ,
\end{eqnarray}
it follows that 
\beq
\delta_0^{(0)}  =  -{i\over 2} 
\log\left[ 1 + i {p M\over 2\pi} \ \Czzm\right]
\ ,\qquad 
\delta_0^{(1)} = {p M\over 4\pi} 
{ \Czzz\over   1 + i {p M\over 2\pi} \ \Czzm }\ ,
\eeq
\beq
\varepsilon_{1}^{(0)}  =  0
\ ,\qquad
\varepsilon_{1}^{(1)} = {p M\over 4\pi} { \Cztz   \over
\sqrt{1 + i {p M\over 2\pi} \ \Czzm } }\ ,
\eeq
\beq
\delta_{2}^{(0)} =  0
\ ,\qquad
\delta_{2}^{(1)} = {p M\over 4\pi} \ \Cttz 
\  .
\eeq

Working to subleading $O(Q^0)$ there are three parameters describing the 
$\siii$ phase shift. Again it is convenient to choose the subtraction point
equal to $m_{\pi}$.
However, as we discussed previously observables do not depend upon $C_0^{(\siii)}(m_{\pi})$ and 
$D_2^{(\siii)}(m_{\pi})$ independently. 
Therefore we can set $D_2(m_\pi)=0$ and fit $C_0$ and  
$C_2$ to the low energy observables taken to be the  scattering length 
$a_3=5.423 \pm 0.005\ {\rm fm}$ and the 
deuteron binding energy ${\it E}_D = 2.224644\pm 0.000034\ {\rm MeV}$.
The result of this fit is
\beq
C^{(\siii)}_0(m_{\pi})=-5.03~{\rm fm}^2\ ,\qquad D^{(\siii)}_2(m_{\pi})\equiv 0\ ,
\qquad C^{(\siii)}_2(m_{\pi})=4.96~{\rm fm}^4\ .
\eeq

\begin{figure}[t]
\centerline{\epsfysize=4 in \epsfbox{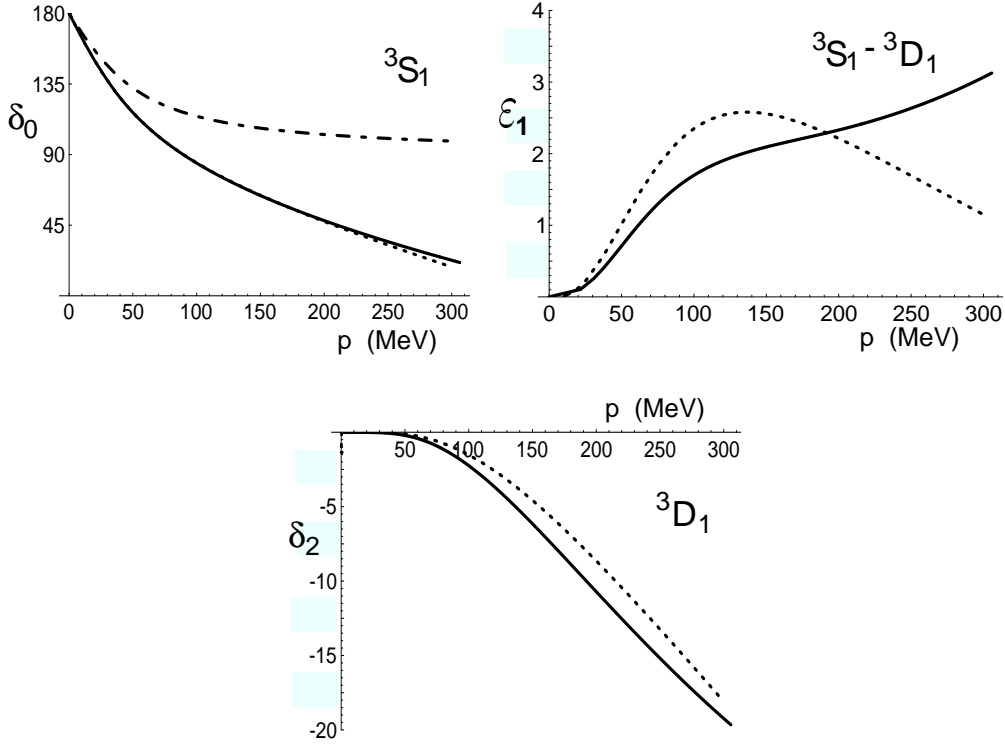}}
\noindent
\caption{\it The phase shifts $\delta_0$, $\delta_2$ and  mixing parameter
$\varepsilon_1$ for the $\siii-\diii$ channel. The solid line denotes the results of the 
Nijmegen partial wave  analysis.
The dot-dashed curve is the fit at $O(Q^{-1})$ for $\delta_0$, while
$\delta_2 = \varepsilon_1 = 0$ at this order.
The dashed curves are the results of the $O(Q^{0})$  fit of $\delta_0$ 
to the partial wave analysis over the momentum range $p\le 200\ \MeV$,  as given in \eq{numfitc}.
}
\label{tripfit}
\vskip .2in
\end{figure}

Alternatively fitting the parameters $C_0^{(\siii)}(m_{\pi})$, 
$C_2^{(\siii)}(m_{\pi})$
 and  $D_2^{(\siii)}(m_{\pi})$  to the phase shift $\delta_0$
over the momentum range $p\le 200 \MeV$ yields
\begin{equation}
C_0^{(\siii)}(m_{\pi})=-5.51\fm^2\ ,\qquad
D_2^{(\siii)}(m_{\pi})=1.32\fm^4\ ,\qquad
C_2^{(\siii)}(m_{\pi})=9.91\fm^4\ 
\  .
\eqn{numfitc}
\end{equation}
 Fig.~\ref{tripfit} 
shows the 
phase shift compared to the 
Nijmegen partial wave  analysis \cite{nijmegen} for this latter set of
parameters.
The other two quantities, $\varepsilon_1 $ and $\delta_2 $,
receive no leading order contributions and both
begin at $O(Q^0)$.
There are no free parameters at this order 
in either $\varepsilon_1 $ or $\delta_2 $
once $C_0^{(\siii)}$
has been determined from $\delta_0$.
The predictions for $\varepsilon_1 $ and $\delta_2 $ from the fit \eq{numfitc} and a comparison
to the Nijmegen partial wave analysis can be found in Fig.~\ref{tripfit}.
%

\subsection{Higher partial waves}

The analysis of the previous section demonstrates how
the power counting impacts the $\diii$ channel and 
this discussion generalizes to other partial waves.
A local operator that connects an angular momentum $L$ state
with an angular momentum $L^\prime$ state involves at least
$L+L^\prime$ spatial derivatives.
The case of $S$-wave to $S$-wave scattering 
has been described 
in the previous sections.
If either $L$ or $L^\prime$ but not both correspond to an $S$-wave 
then 
the operator enters at $O(Q^{L+L^\prime-1})$.
However, if neither $L$ nor $L^\prime$ is equal to zero the operator
contributes at $O(Q^{L+L^\prime})$ for $L, L^\prime$ odd and 
$O(Q^{L+L^\prime-1})$ for $L, L^\prime$ even. 
The contribution of pions is at $O(Q^0)$, and is therefore
the leading contribution to all non $S$-wave to $S$-wave 
scattering amplitudes. 
This contribution has been presented  in the 
literature (e.g. \cite{EricW}).

\subsection{Radiation pions and operator mixing}

We have seen that graphs involving potential pion exchange occur at $O(Q^0)$ 
 in both the $\si$ and $\siii-\diii$ channels.
Such contributions arise from kinematic regions where the intermediate 
nucleons are near their mass-shell, with the pion exchanged 
in the $t$-channel far from its pole.
Contributions arising from radiation pions  (the pion pole)
exchanged in the $s,t$ and $u$ channels arise at $O(Q^1)$
(for discussion of radiation exchanges in nonrelativistic gauge theories
see 
\cite{LukeManohar2,Labelle,GrinsteinRothstein,LukeSavage,Beneke,Griesshammer}).  
An interesting
feature of virtual radiation pions in the $s$-channel 
is that they can cause mixing between four-nucleon operators
in different spin channels.  This is because the two nucleons in a virtual
intermediate $NN\pi$ state can rescatter while
in a different isospin and angular momentum state than the physical incoming nucleon pair.

\begin{figure}[t]
\centerline{\epsfysize=1 in \epsfbox{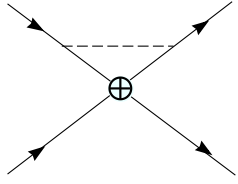}}
\noindent
\caption{\it A graph involving a radiation pion that gives rise to mixing between the 
$\si-\siii$ operators. This graph  with an insertion of $(N^\dagger\sigma^a N)^2$ at 
the vertex gives the expression $\Gamma_1$ in  \eqs{gama}{gamc}.}
\vskip .2in
\label{radpi}
\end{figure}
As an  example, consider the graph shown in Fig.~\ref{radpi}
with an insertion of  the operator $(N^\dagger \sigma^a N)^2 $, denoted by $\Gamma_1$.
Explicitly, 
\begin{eqnarray}
\Gamma_1 & = &  i  {g_A^2\over 2 f^2}
N^\dagger\tau^a\sigma^i\sigma^k N  
N^\dagger \tau^a\sigma^k \sigma^j N
 \nonumber\\
& & 
\left({\mu\over 2}\right)^{4-D}
\int {d^D q\over (2\pi)^D} { q^i q^j \over 
\left[ q^0 + {E\over 2} - {({\bf p + q})^2\over (2 M)} + i \epsilon\right]
\left[ q^0 + {E\over 2} - {({\bf p^\prime + q})^2\over (2 M)} + i \epsilon\right]
\left[ (q^0)^2 - {\bf q}^2 - m_\pi^2 + i \epsilon\right]}
\ .
\eqn{gama}
\end{eqnarray}
The $q^0$ integral is performed by forming a contour enclosing the one
pole in the 
upper half of the complex plane provided by the pion propagator.
Using the equations of motion 
$E={\bf p}^2/M + O({\bf p}^4/M^3)$ and neglecting the 
$O({\bf p}^4/M^3)$ relativistic correction
we realize that the weight of the integral in the low energy theory 
will be for momentum near the pion mass and perform an expansion in $1/M$,
giving
\begin{eqnarray}
\Gamma_1 & = &  {g_A^2\over 4 f^2} 
N^\dagger\tau^a\sigma^i\sigma^k N  
N^\dagger \tau^a\sigma^k \sigma^j N
\left({\mu\over 2}\right)^{4-D}
\int {d^{D-1} q\over (2\pi)^{D-1}} { q^i q^j \over 
\left[{\bf q}^2 + m_\pi^2\right]^{3\over 2}  }
\ +\ O(1/M)
\ .
\eqn{gamb}
\end{eqnarray}
Evaluating the integral 
yields
\begin{eqnarray}
\Gamma_1 & = &  
{3 g_A^2 m_\pi^2 \over 32\pi^2  f^2}  
\left[ {1 \over \epsilon} - \log\left( {m_\pi^2\over\mu^2}\right) + {\rm constant} \right] 
\left[  \(N^\dagger \sigma^a N\)^2 
\ +\ \(N^\dagger  N\)^2  \right]
\ +\ O(1/M)
\ .
\eqn{gamc}
\end{eqnarray}

Including all the irreducible graphs and wavefunction renormalization 
it is straightforward to find the 
leading radiation pion contribution to the $\beta$ functions 
for $D_2^{(\si)}$ and $D_2^{(\siii)}$ are
\begin{eqnarray}
\beta_{D_2^{(\si)}}^{(rad)} & = & +\ {3 g_A^2  
\over 4 \pi^2 f^2} \left( C_0^{(\siii)} - C_0^{(\si)} \right)
\ \ \ ,
\nonumber\\
\beta_{D_2^{(\siii)}}^{(rad)} & = & -\ {3 g_A^2  
\over 4 \pi^2 f^2} \left( C_0^{(\siii)} - C_0^{(\si)} \right)
\ .
\end{eqnarray}
These contributions give rise to mixing between the $S$-wave 
spin-singlet and spin-triplet operators.

\subsection{Relativistic effects}

\begin{figure}[t]
\centerline{\epsfysize=2 in \epsfbox{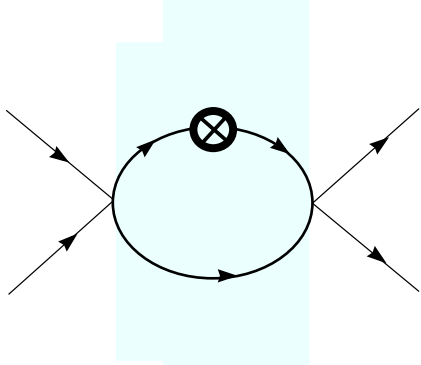}}
\noindent
\caption{\it Higher dimension operators arising from relativistic corrections
to the energy-momentum relation, denoted by the crossed-circle,  
are inserted perturbatively. }
\vskip .2in
\label{pfour}
\end{figure}

A further contribution starting at $O(Q^1)$ arises from 
relativistic corrections
to the energy-momentum relation.
A detailed discussion of such effects in dimensionally regulated 
non-relativistic gauge theories can be 
found in \cite{LukeSavage} and the situation is similar for 
nucleon interactions.
Neglecting pion fields, 
the lagrange density in the single nucleon sector is
\begin{eqnarray}
{\cal L} & = & 
N^\dagger\left( i\partial_0 + {\nabla^2\over 2 M}\right) N
\ +\ 
{1\over 8 M^3} N^\dagger \nabla^4 N
\ +\ ...
\ ,
\end{eqnarray}
where the spatial gradient operator brings down factors of 
${\bf p}$.
It is understood that the  $N^\dagger \nabla^4 N$ operator is 
inserted perturbatively into graphs, e.g. Fig.~\ref{pfour},
and that the lowest order
equations of motion are modified to 
$E/2 = {\bf p}^2/2 M + {\bf p}^4/ 8 M^3 +\ldots$.
A single insertion of $N^\dagger \nabla^4 N$ gives rise to a 
$O(Q^1)$ contribution; however, it is suppressed by factors of the 
nucleon mass and not $\Lambda_{NN}$.
Consequently, its effect is expected to be  small
compared to other corrections at this order.
%

\section{Overview}

In a  previous letter \cite{KSWb}\ 
we presented a new power counting scheme to describe 
NN scattering  processes
that does not suffer from the inconsistencies 
of Weinberg's scheme \cite{Weinberg1}.
In order to achieve consistent power counting a  new subtraction scheme
is used for dimensionally regulated integrals, designed to keep track of power law divergences
in theories in which there are delicate cancellations at short distance; for $NN$ scattering,
these cancellations manifest themselves as a large scattering length.
The renormalization group provides a  powerful tool in the 
analysis of such theories and allows one to identify the order 
at which a given operator will contribute.

In the present paper we have  elaborated on our expansion, with particular emphasis on the utility of 
the renormalization group.  We have also presented detailed analytic computations 
for $NN$ scattering in the $\si$ and $\siii-\diii$ channels, complete to the  subleading $O(Q^0)$ order,
and have shown that the agreement with experiment at low energy is quite good.  A particular success was
the  calculation of the $\siii-\diii$ mixing parameter $\varepsilon_1$ shown in Fig.~\ref{tripfit}, with
no free parameters.  
We then discussed the power counting for higher partial waves, and 
showed that they are dominated by pion exchange in the Born approximation. 
Finally, we showed how to deal with  relativistic effects and virtual radiative pions, which are 
 features that arise at order  $O(Q^1)$.

The techniques presented here should be applicable to a number of low energy processes, such as radiative
capture, and electromagnetic moments of the deuteron.  Theoretical challenges include extending 
the validity of the expansion above the scale $\Lambda_{NN}$, and applying the technique to systems with
three or more nucleons.

\bigskip\bigskip


We would like to thank G. Bertsch  for
useful discussions.
This work is supported in part by the U.S. Dept. of Energy under
Grants No. DOE-ER-40561,  DE-FG03-97ER4014, and DE-FG03-92-ER40701.



\begin{references}

\bibitem{ManRev} A.V. Manohar,  
Lectures given at 35th Internationale Universitatswochen fuer Kern- und Teilchenphysik,
{\it Perturbative and Nonperturbative Aspects of Quantum Field Theory}, 
Schladming, Austria, 2-9 Mar 1996. 
{\tt hep-ph/9606222 }.

\bibitem{KSWb} D.B. Kaplan, M.J. Savage and M.B. Wise, 
{\tt nucl-th/9801034}, {\it to appear in Phys. Lett. B}.

\bibitem{Weinberg1}S. Weinberg,
Phys. Lett. {\bf B251} (1990) 288;
Nucl. Phys. {\bf B363} (1991) 3;
Phys. Lett. {\bf B295} (1992) 114.


\bibitem{KoMany} C. Ordonez and U. van Kolck, Phys. Lett. {\bf B291} (1992) 459;
C. Ordonez, L. Ray and  U. van Kolck, Phys. Rev. Lett. {\bf 72} (1994) 1982;
Phys. Rev. {\bf C53} (1996) 2086.;  
U. van Kolck, Phys. Rev. {\bf C49} (1994) 2932.

\bibitem{Parka} T.S.  Park, D.P.  Min and M. Rho,
Phys. Rev. Lett. {\bf 74} (1995) 4153;
Nucl. Phys. {\bf A596} (1996) 515.

\bibitem{KSWa} D.B. Kaplan, M.J. Savage and M.B. Wise,
Nucl. Phys. {\bf B478} (1996) 629, {\tt nucl-th/9605002}.


\bibitem{CoKoM} T. Cohen, J.L. Friar, G.A. Miller and 
U. van Kolck, Phys. Rev. {\bf C53} (1996), 2661.

\bibitem{DBK} D. B. Kaplan, Nucl. Phys. {\bf B 494} (1997) 471.

\bibitem{cohena}T.D. Cohen, Phys. Rev. {\bf C55} (1997)  67.
D.R. Phillips and T.D. Cohen, Phys. Lett. {\bf B390} (1997) 
7.  
K.A. Scaldeferri, D.R. Phillips, C.W. Kao and T.D. Cohen,
Phys. Rev. {\bf C56} (1997) 679.
S.R. Beane, T.D. Cohen and D.R. Phillips,
nucl-th/9709062.
 

\bibitem{Fria} J.L. Friar,  Few Body Syst. {\bf 99} (1996) 1,
{\tt nucl-th/9607020}. 

\bibitem{Sa96} M.J. Savage, Phys. Rev. {\bf C55} (1997) 2185,
{\tt nucl-th/9611022}. 

\bibitem{LMa} M. Luke and A.V. Manohar, 
Phys. Rev. {\bf D55} (1997)  4129,  
{\tt hep-ph/9610534 }.

\bibitem{GPLa} G.P. Lepage, {\tt nucl-th/9706029},
Lectures given at 9th Jorge Andre Swieca Summer School: 
Particles and Fields, Sao Paulo,
Brazil, 16-28 Feb 1997.

\bibitem{Adhik} S.K. Adhikari and A. Ghosh, 
J. Phys. {\bf A30} (1997) 6553.

\bibitem{RBMa}  K.G. Richardson, M.C. Birse and J.A. McGovern,
{\tt hep-ph/9708435}.
 
\bibitem{Bvk} P.F. Bedaque and U. van Kolck, 
{\tt nucl-th/9710073};
P.F. Bedaque, H.-W. Hammer and U. van Kolck,
{\tt nucl-th/9802057}.


\bibitem{Parkb} T.S.  Park, K. Kubodera,  D.P.  Min and M. Rho, 
{\tt hep-ph/9711463}.


\bibitem{SavageWise} M. J. Savage and M. B. Wise,
    Phys. Rev. {\bf D53} (1996),
    {\tt hep-ph/9507288}.


\bibitem{KaplanNelson}    D.B. Kaplan and A.E. Nelson,
    Phys. Lett. {\bf B175} ( 1986) 57;
    A. E. Nelson and D. B. Kaplan,
    Phys. Lett. {\bf 192B} (1987) 193.

\bibitem{kcon} H.D. Politzer and M.B. Wise, Phys. Lett. {\bf B257} (1991) 399;
G.E. Brown, C.-H. Lee, M. Rho, V. Thorsson, Nucl. Phys. {\bf A567} (1994) 937;
C.M. Ko, V. Koch and G. Li,
{\tt nucl-th/9702016};
G.Q. Li, C.H. Lee and G.E. Brown,
Nucl. Phys. {\bf A625} (1997) 372, 
{\tt nucl-th/9706057}.

\bibitem{Gegelia} J. Gegelia,
{\tt  nucl-th/9802038}.

\bibitem{Weinberg2} S. Weinberg,
Phys. Rev. Lett. {\bf 17} (1966) 616;
Phys. Rev. {\bf 166}  (1968) 1568.

\bibitem{GeorgiManohar}    A. Manohar and  H. Georgi,
Nucl. Phys. {\bf B234} (1984) 189.

\bibitem{GasserLeutwyler}J. Gasser and H. Leutwyler,  
Annals Phys. {\bf 158} (1984) 142;
Nucl. Phys. {\bf B250} (1985) 465.  

\bibitem{Kaplan2} A. G. Cohen, D.B. Kaplan and A.E. Nelson,  
Phys. Lett. {\bf B412} 301,
{\tt  hep-ph/9706275}

\bibitem{LukeManohar2}     M. Luke and A.V. Manohar,
Phys. Lett. {\bf B286} (1992) 348,
{\tt hep-ph/9205228}

\bibitem{Labelle}  P. Labelle, 
{\tt hep-ph/9608491}

\bibitem{GrinsteinRothstein} B. Grinstein and I.Z. Rothstein, 
Phys. Rev. {\bf D57} (1998) 78.
{\tt hep-ph/9703298}

\bibitem{LukeSavage}  M. Luke and M.J. Savage,
Phys. Rev. {\bf D57} (1998)  413.
{\tt hep-ph/9707313}.


\bibitem{Beneke} M. Beneke and V.A. Smirnov, 
{\tt  hep-ph/9711391}

\bibitem{Griesshammer} H.W. Griesshammer,
{\tt  hep-ph/9712467}.

\bibitem{nijmegen} 
V.G.J. Stoks, R.A.M. Klomp, C.P.F. Terheggen and J.J. de Swart,
Phys. Rev. {\bf C49} (1994) 2950,
{\tt nucl-th/9406039}.

\bibitem{EricW} {\it Pions and Nuclei} by T. Ericson and W. Weise,
Oxford Science Publications (1988); ISBN 0-19-852008-5.


\end{references}
\end{document}